\documentclass[10pt, conference, letterpaper]{IEEEtran}
\usepackage[top=0.75in, bottom=1.1in, left=.75in, right=.75in]{geometry}
\usepackage{balance}
\usepackage{cite}
\usepackage{graphicx}
\usepackage{subcaption} 
\usepackage{epsfig}
\usepackage{epstopdf}
\usepackage{enumitem}
\usepackage{wrapfig}
\usepackage{amsmath}
\usepackage{amsthm}
\usepackage{url}
\usepackage{graphicx}
\usepackage{booktabs} 
\usepackage{siunitx} 

\usepackage{algorithm}
\usepackage{algorithmic}
\usepackage{bbm}
\usepackage{amsfonts}

\usepackage{xcolor}
\ifCLASSINFOpdf
\else
\fi
\usepackage{multirow}

\usepackage{colortbl}

\definecolor{darkblue}{rgb}{0.0, 0.0, 0.55}

\begin{document}

\title{ Demo: Secure Edge Server for Network Slicing and Resource Allocation in Open RAN}
\author{\textbf{Adhwaa Alchaab$^*$, Ayman Younis, Dario Pompili$^*$}\\{ $*$Department of Electrical and Computer Engineering}\\
{$*$Rutgers University--New Brunswick, NJ, USA}\\
{E-mails: \{adhwaa.alchaab, a.younis, pompili\}@rutgers.edu
}}
\maketitle

\begin{abstract} 
Next-Generation Radio Access Networks (NG-RAN) aim to support diverse vertical applications with strict security, latency, and Service-Level Agreement (SLA) requirements. These demands introduce challenges in securing the infrastructure, allocating resources dynamically, and enabling real-time reconfiguration. This demo presents SnSRIC, a secure and intelligent network slicing framework that mitigates  a range of Distributed Denial-of-Service (DDoS) attacks in Open RAN environments. SnSRIC incorporates an AI-driven xApp that dynamically allocates Physical Resource Blocks (PRBs) to active users while enforcing slice-level security. The system detects anomalous behavior, distinguishes between benign and malicious devices, and uses the E2 interface to throttle rogue signaling while maintaining service continuity for legitimate users.

\end{abstract}

\section {Introduction}

Recent advancements in AI-based network slicing have shown promising results in dynamically managing radio resources in terms of different time scales to improve Quality of Service (QoS) in O-RAN environments~\cite{10558834,8793038}. To achieve the vision of Next-G RAN, which includes high data throughput and ultra-low latency along with real-time edge cloud processing, the O-RAN framework has emerged as a cornerstone of AI-enabled 6G mobile networks, representing a transformative shift in RAN architecture~\cite{oran_wg1}.
It extends the 3GPP-defined architecture by introducing the RAN Intelligent Controller (RIC), which enables networks to harness AI capabilities through open interfaces. Network slicing is a prominent feature of O-RAN that provides end-to-end connectivity 
tailored to specific service requirements such as the support for very high data rates, 
traffic densities or ultra-low latency based on a SLA~\cite{10723502}.

Despite its benefits, the open and flexible design of O-RAN introduces significant security challenges. Traditional security frameworks, designed for less dynamic networks, are inadequate for O-RAN’s unique environment. For example, mobile device design and regulation typically impose limits on the number of signaling events a device can generate, keeping the average volume relatively low. However, some devices occasionally violate these restrictions. Attackers exploit such devices by recruiting them into botnets to launch multiple signaling storms, resulting in DDoS attacks on the RAN. Currently, there are no effective controls within the RAN to mitigate these coordinated attacks, and signaling flow throttling in the core network lacks the ability to distinguish between benign and malicious devices~\cite{10364950}. Ideally, such attacks should be blocked as close to the source as possible—at the network edge. Without proper safeguards, the network remains vulnerable to threats such as malicious traffic, DDoS attacks, jamming, and data channel intrusions. Moreover, the use of commercial off-the-shelf software-defined radios (SDRs) further heightens these risks by enabling attacks such as fake base stations and malicious user equipment. \emph{This demo presents SnSRIC, a novel secure network slicing framework in O-RAN, showcasing how network slices can be dynamically managed to satisfy SLA requirements while maintaining security for each slice by detecting and mitigating diverse threats at the RAN edge. It highlights practical, AI-driven solutions that balance both performance and security, advancing the design of robust and resilient next-generation networks.}

\begin{figure*}[!ht]
 \centering
 \begin{tabular}{ccc}
\hspace*{-.3cm}\includegraphics[width=0.28\textwidth]{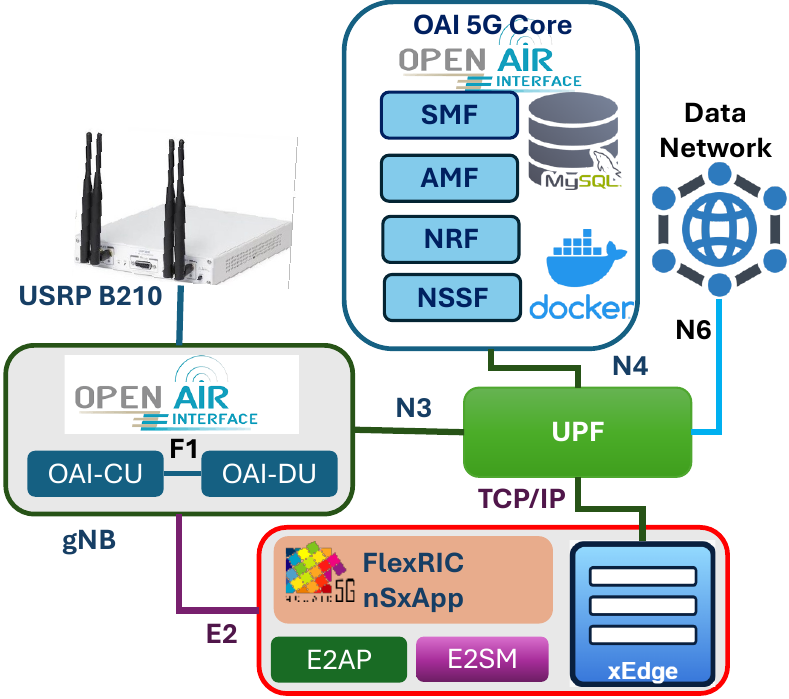}&
\hspace*{-.1cm}\includegraphics[width=0.32\textwidth]{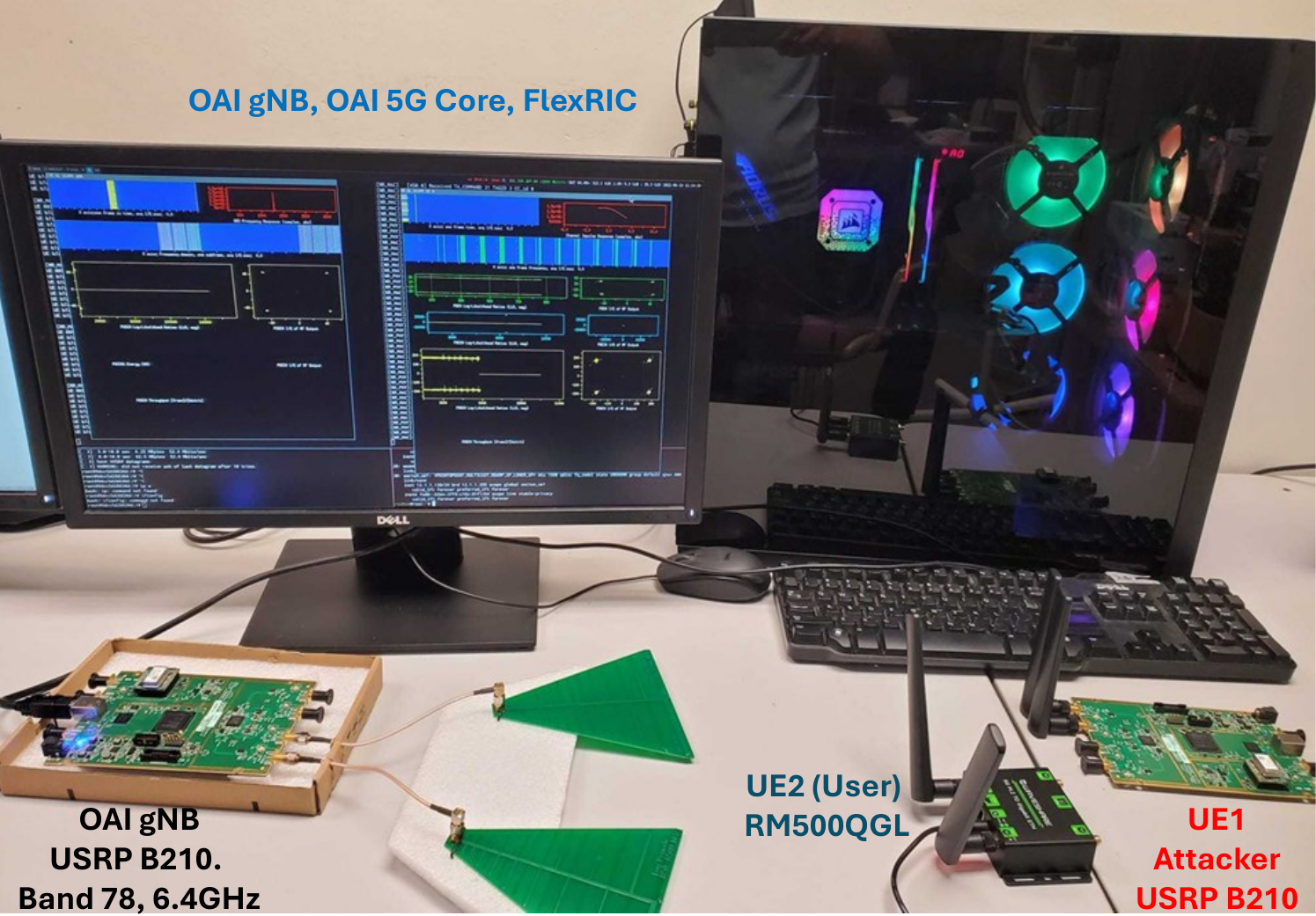} &
\hspace*{-.1cm}\includegraphics[width=0.34\textwidth]{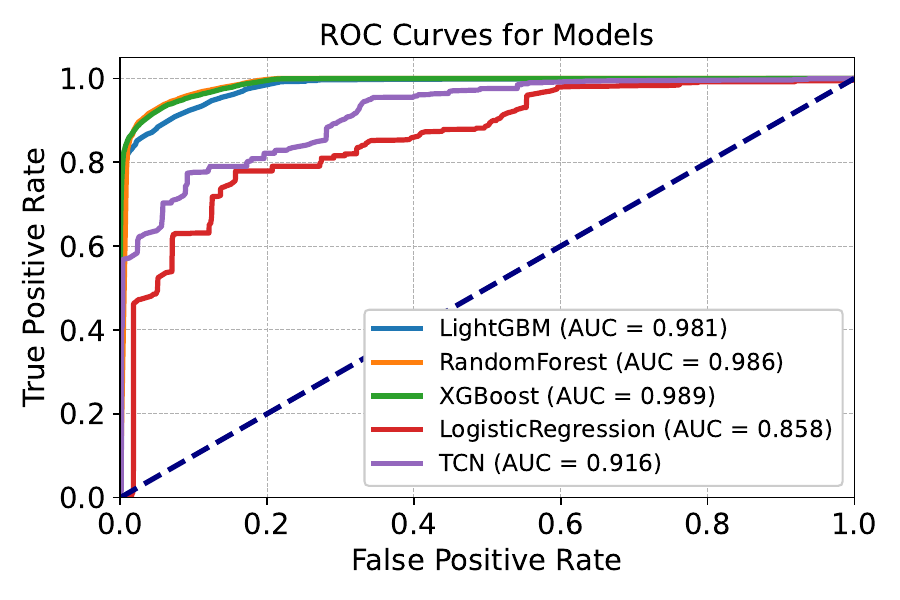} \\
\small (a) & \small(b) & \small(c)
\end{tabular}
\caption{(a) Logical testbed diagram;(b) SDR-based testbed setup; (c) ROC curves of AI-based anomaly detectors; }\label{fig:SL}
\vspace{-0.13in}
\end{figure*}

\begin{figure*}[t]
    \centering
    \begin{minipage}[b]{0.32\textwidth}
        \centering
        \includegraphics[width=\textwidth]{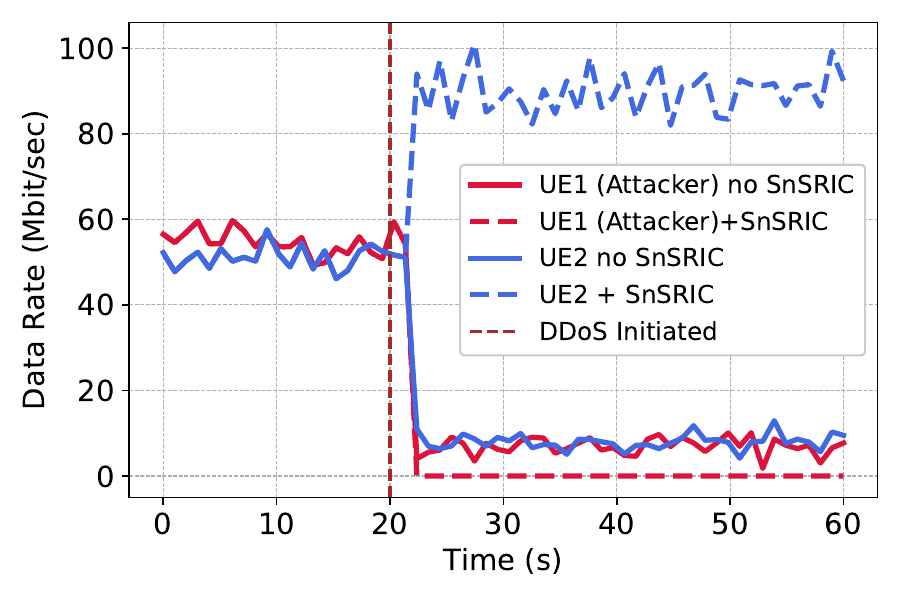}
        \small (a)
    \end{minipage}
    \hfill
    \begin{minipage}[b]{0.32\textwidth}
        \centering
        \includegraphics[width=\textwidth]{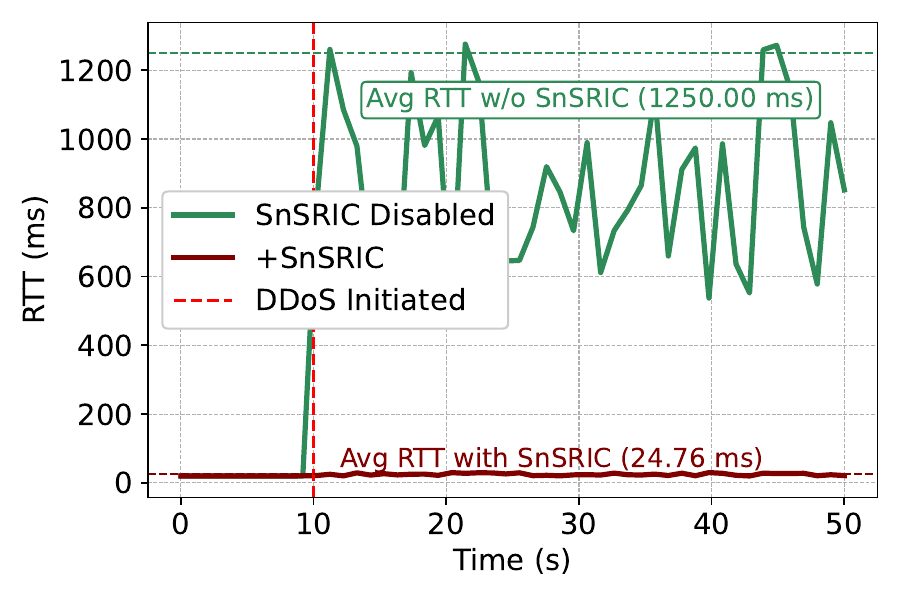}
        \small (b)
    \end{minipage}
    \hfill
    \begin{minipage}[b]{0.32\textwidth}
        \centering
        \includegraphics[width=\textwidth]{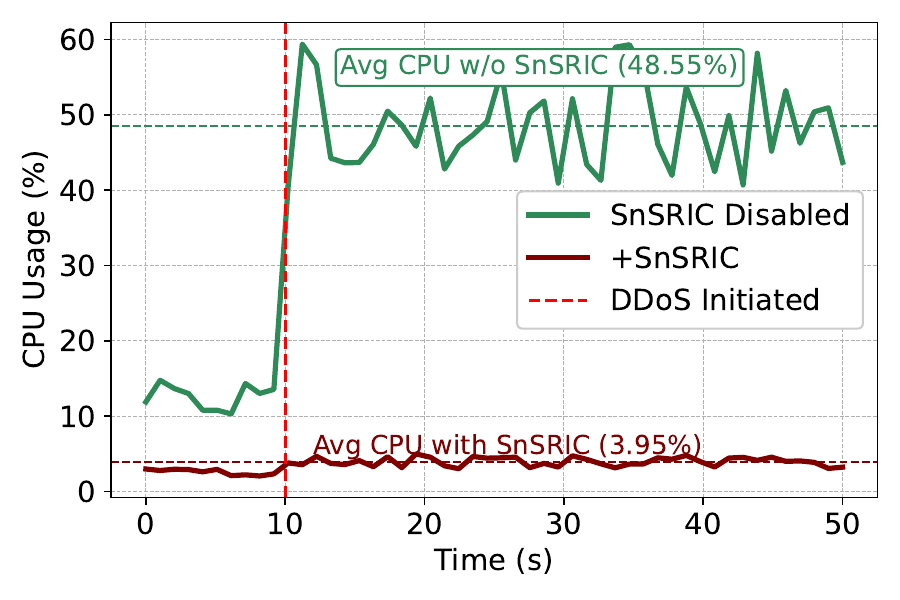}
        \small (c)
    \end{minipage}
    \vspace{-0.1in}
    \caption{(a) UE1/UE2 data rate during DDoS under baseline and SnSRIC  (b) RTT improvements during an attack scenario; and (c) CPU usage in normal vs. attack conditions.}
    \label{fig:ro}
    \vspace{-0.2in}
\end{figure*}

 \section{SnSRIC Integration}
 The O-RAN architecture is well described in~\cite{polese2023understanding}. Therefore, this section focuses on the proposed framework, SnSRIC (see Fig.~\ref{fig:SL}(a)), which is integrated into FlexRIC and OpenAirInterface~(OAI)~\cite{openairinterface}. The architecture consists of two primary components: (i) the nSxApp, an AI-driven xApp deployed in the Near-Real-Time RIC (Near-
RT RIC) to manage network slicing; and (ii) a distributed, per-slice xEdge server that enables anomaly detection and threat mitigation within the RAN.

\textbf{AI-Based SLA Network Slicing xApp (nSxApp).} We first present the \textit{nSxApp}, an AI-driven network slicing xApp that operates within the Near-RT RIC to dynamically manage radio resource allocation in response to evolving traffic patterns and service-level requirements. It handles slice instantiation, scheduling policy enforcement, and UE-to-slice associations through standardized E2 Service Models, specifically E2SM Slice Control and Slice Indication. To support differentiated service delivery across network slices, the nSxApp implements a suite of slice scheduling algorithms, including:
\emph{i)~STATIC}, which assigns fixed and non-overlapping sets of PRBs to each slice, ensuring strict service isolation and predictable performance; \emph{ii)~Network Virtualization Substrate (NVS)}, which enables elastic resource provisioning through both rate-based and capacity-based reservation schemes. The rate-based mode configures minimum guaranteed and reference throughput levels, while the capacity-based mode proportionally allocates resources based on reserved percentage shares of total PRBs; \emph{iii)~Earliest Deadline First (EDF)}, which prioritizes resource allocation based on application-level time constraints, making it particularly effective for latency-sensitive services such as augmented reality, remote control, and 
industrial automation. The nSxApp interfaces with the OAI gNB via the E2 interface, supporting both monolithic and disaggregated Central Unit and Distributed Unit (CU and DU) configurations. To enable intelligent threat-aware resource management, the architecture incorporates a distributed per-slice \textit{xEdge} server positioned at the network edge. This server performs continuous traffic monitoring and computes anomaly scores based on protocol-level features. These scores are transmitted to the nSxApp over a TCP socket connection, enabling it to perform fine-grained resource control for each user equipment. This includes dynamically adjusting the allocation of PRBs and triggering Radio Resource Control connection release (RRC) procedures for suspicious users. This coordinated approach ensures that radio resource decisions remain consistent with service level agreement requirements while maintaining a high QoS for end users.

\textbf{xEdge Server.}
The \textit{xEdge} server, co-located with the User Plane Function (UPF), functions  as a lightweight security layer designed to enable proximal threat detection and policy enforcement at the network edge. It integrates an AI-based anomaly detection module powered by a pre-trained LightGBM classifier, selected based on its efficient trade-off between inference time, memory utilization, and detection accuracy. As shown in Table~\ref{tab:model_comparison} and  Fig.~\ref{fig:SL}(c), LightGBM consistently shows efficient performance when evaluated against several baseline models. The LightGBM model was trained on two benchmark datasets, KDDCup'99~\cite{kddcup99} and UNSW-NB15~\cite{unsw-nb15}. 
From each observed packet, the system extracts a feature vector comprising: (i) \texttt{protocol\_type}, identifying the transport layer protocol (e.g., TCP, UDP, ICMP); (ii) \texttt{service}, inferred from the destination port; (iii) \texttt{flag}, encoding TCP control flags (e.g., SYN, ACK); (iv) \texttt{src\_bytes}, representing the payload size sent by the source; and (v) \texttt{dst\_bytes}, denoting the payload size received by the destination. These features are accumulated in a sliding window of $40~\rm{packets}$, preprocessed, and passed to the LightGBM model for inference. The model classifies each packet as benign or anomalous and generates a summary report, which is transmitted via TCP socket to the nSxApp. This enables the nSxApp to enforce adaptive, slice-aware responses to network threats. Based on anomaly scores from the xEdge server, it proportionally throttles PRB allocations. If a user reaches $100\%$ anomaly, the nSxApp triggers an RRC release to disconnect them.

\begin{table}[!ht]
\centering
\caption{Model comparison across various datasets.}
\label{tab:model_comparison}
\vspace{-0.06in}
\footnotesize
\begin{tabular}{|l|c|c||c|c|}
\hline
\multirow{2}{*}{\textbf{Model}} & \multicolumn{2}{c||}{\textbf{UNSW-NB15}} & \multicolumn{2}{c|}{\textbf{KDDCup'99}} \\
\cline{2-5}
 & \textbf{Acc.} & \textbf{Inf. (ms)} & \textbf{Acc.} & \textbf{Inf. (ms)} \\
\hline
LightGBM            & 0.93  & 3.00   & 0.999 & 5.00 \\
Random Forest       & 0.94  & 171.5  & 1.000 & 187.2 \\
XGBoost             & 0.94  & 4.00   & 0.999 & 9.51 \\
Log. Regression     & 0.75  & 1.00   & 0.954 & 6.12 \\
TCN                 & 0.81  & 2.98   & 0.987 & 4.00 \\
\hline
\end{tabular}
\vspace{-0.1in}
\end{table}

\section{ Deployment and  Evaluation}
This section describes the SnSRIC deployment on a standalone 5G SDR testbed and evaluates its performance across slicing and anomaly detection scenarios.

\textbf{Demo Setup.} 
 To evaluate our proposed SnSRIC framework in real-world scenario, we integrated this experimental testbed, as illustrated in Fig.~\ref{fig:SL}(b), which comprises the following main components: \emph{i)~gNB:} The gNB is implemented using the OAI 5G stack, including the CU and DU, which are deployed on a high-performance server. The CU/DU are connected to a USRP B210 that functions as the Radio Unit (RU), enabling over-the-air transmission. 
\emph{ii)~5G Core Network:} The 5G core is also implemented using OAI, providing all key components required for a complete  5G deployment—including the Access and Mobility Management Function (AMF), Session Management Function (SMF), UPF, and other control plane elements. This setup supports data routing, authentication, session handling, and mobility management for connected UEs.
\emph{iii)~O-RAN RIC:}  We utilize FlexRIC, a flexible and modular RIC platform that supports near-real-time functionalities. The Near-RT RIC handles network slicing and resource management via nSxApp. \emph{iv)~5G UEs:} The UE side consists of multiple endpoints. We utilize OAI 5G UE software running on a USRP B210, as well as a Quectel RM500Q-GL 5G modem. 

\textbf{Experimental Results.} 
We evaluated SnSRIC on a 5G SDR testbed under various scenarios, focusing on resource usage, latency, and CPU overhead.

\emph{Baseline Behavior.}
Two UEs were connected to distinct network slices under a static scheduling configuration. Both UEs transmitted legitimate traffic, enabling the nSxApp to maintain fair PRB allocation. As shown in Fig.~\ref{fig:ro}(a), both users maintained stable and nearly equal data rates ($\sim$55--60~\text{Mbit/s}), indicating balanced PRB allocation. The average RTT in Fig.~\ref{fig:ro}(b) remained below $30~\mathrm{ms}$, and CPU utilization  at the core in Fig.~\ref{fig:ro}(c) stayed under $4\%$. Since no anomalies were present, the system operated under normal conditions, serving as a baseline for evaluating system behavior prior to any malicious activity.

\emph{Dynamic Resource Allocation.} To evaluate system behavior under dynamic scheduling, the \texttt{NVS} algorithm was applied, allowing PRB allocation to adapt based on throughput demand. In this case, UE1 injected adversarial traffic by replaying records from the KDDCup’99 and UNSW-NB15 datasets, while UE2 continued normal operation. Without SnSRIC, the nSxApp lacked awareness of the nature of the traffic, which led to an unfair allocation of PRBs in favor of the malicious UE. As shown in Fig.~\ref{fig:ro}(a), the downlink data rate for UE2 decreased sharply after the attack began, while UE1 continued to consume significant bandwidth, which highlights the failure to isolate the malicious user. RTT in Fig.~\ref{fig:ro}(b) increased to an average of $1250~\mathrm{ms}$, and CPU usage in Fig.~\ref{fig:ro}(c) averaged $48.55\%$, indicating high resource consumption and instability under attack conditions.

\emph{Threat-Aware Resource Control.} Once the SnSRIC framework was deployed, the xEdge server continuously analyzed traffic using a pretrained LightGBM-based anomaly detector, which identified abnormal patterns from UE1 and reported them to the nSxApp. The nSxApp responded by throttling PRBs allocated to UE1 and subsequently releasing its RRC connection. As a result, UE2’s throughput in Fig.~\ref{fig:ro}(a) recovered, RTT in Fig.~\ref{fig:ro}(b) returned to sub-$30~\mathrm{ms}$ levels, and CPU usage in Fig.~\ref{fig:ro}(c) averaged below $5\%$. These metrics confirm the efficacy of SnSRIC in isolating the attacker and reallocating resources to preserve QoS for compliant users.

\begin{table}[t]
\centering
\caption{Experimental setup.}
\vspace{-0.06in}
\begin{tabular}{|l|l||l|l|}
\hline
\textbf{CPU} & Intel i7 2.10 GHz & \textbf{RAN } & OAI, USRB B210 \\
\hline
\textbf{Band} & 78 @ 6.4 GHz & \textbf{RIC} & FlexRIC \\
\hline
\textbf{GPU} &RTX 4090 & \textbf{UE1} & Quectel RM520Q-GL \\
\hline
\textbf{RAM} & 32 GB & \textbf{UE2 } & USRP B210 \\
\hline
\textbf{OS} & Ubuntu 22 & \textbf{PktTool} & Scapy\\
\hline
\textbf{Lib.} & PyTorch & \textbf{Datasets} &  KDDCup99, UNSW-NB15 \\
\hline
\end{tabular}
\label{tab:setup}
\vspace{-0.2in}
\end{table}
\balance

\section{Conclusion}
The SnSRIC framework, built and tested on a real-world 5G SDR testbed, demonstrates high performance in secure network slicing and resource allocation for O-RAN. Through integration with OpenAirInterface and FlexRIC, it effectively manages heterogeneous service requirements, allocates resources dynamically, and mitigates DDoS attacks in real time. Experimental results validate its ability to detect anomalous behavior, isolate malicious UEs, and maintain QoS for legitimate users, ensuring resilient and SLA-compliant service delivery in diverse 5G scenarios.

\textbf{Acknowledgment.} This work was supported by the US NSF under Grant No. ECCS-2030101.

\IEEEpeerreviewmaketitle
\bibliographystyle{ieeetr}\small
\bibliography{dref}
\end{document}